\documentclass[aps,prl,superscriptaddress,reprint]{revtex4-1}
\usepackage{amsmath}
\usepackage{amssymb}
\usepackage{graphicx}
\usepackage{xcolor}

\begin{document}

\newcommand{\hkl}[3]{(#1#2#3)}
\newcommand{\bragg}[3]{#1#2#3}

\newlength{\figurewidth}
\setlength{\figurewidth}{0.75\columnwidth}

\graphicspath{{figures/}{/}}

\title{Localized excited charge carriers generate ultrafast inhomogeneous strain in the multiferroic BiFeO$_3$}

\author{Daniel Schick}
\affiliation{Institut f\"ur Physik \& Astronomie, Universit\"at Potsdam, Karl-Liebknecht-Str. 24-25, 14476 Potsdam, Germany}
\author{Marc Herzog}
\affiliation{Institut f\"ur Physik \& Astronomie, Universit\"at Potsdam, Karl-Liebknecht-Str. 24-25, 14476 Potsdam, Germany}
\affiliation{Abteilung Physikalische Chemie, Fritz-Haber-Institut der Max-Planck-Gesellschaft, Faradayweg 4-6, 14195 Berlin, Germany}
\author{Haidan Wen}
\affiliation{X-ray Science Division, Argonne National Laboratory, Argonne, Illinois 60439, USA}
\author{Pice Chen}
\affiliation{Department of Materials Science and Engineering and Materials Science Program, University of Wisconsin-–Madison, Madison, Wisconsin 53706, USA}
\author{Carolina Adamo}
\affiliation{Department of Material Science and Engineering, Cornell University, Ithaca, New York 14853, USA}
\affiliation{Department of Applied Physics, Stanford University, Stanford, CA 94305-4045, USA}
\author{Peter Gaal}
\affiliation{Helmholtz-Zentrum Berlin f\"ur Materialien und Energie GmbH, Wilhelm-Conrad-R\"ontgen Campus, BESSY II, Albert-Einstein-Str. 15, 12489 Berlin, Germany}
\author{Darrell G. Schlom}
\affiliation{Department of Material Science and Engineering, Cornell University, Ithaca, New York 14853, USA}
\affiliation{Kavli Institute at Cornell for Nanoscale Science, Ithaca, New York 14853, USA}
\author{Paul G. Evans}
\affiliation{Department of Materials Science and Engineering and Materials Science Program, University of Wisconsin-–Madison, Madison, Wisconsin 53706, USA}
\author{Yuelin Li}
\affiliation{X-ray Science Division, Argonne National Laboratory, Argonne, Illinois 60439, USA}
\author{Matias Bargheer}
\email{bargheer@uni-potsdam.de}
\homepage{www.udkm.physik.uni-potsdam.de}
\affiliation{Institut f\"ur Physik \& Astronomie, Universit\"at Potsdam, Karl-Liebknecht-Str. 24-25, 14476 Potsdam, Germany}
\affiliation{Helmholtz-Zentrum Berlin f\"ur Materialien und Energie GmbH, Wilhelm-Conrad-R\"ontgen Campus, BESSY II, Albert-Einstein-Str. 15, 12489 Berlin, Germany}

\date{\today}

\begin{abstract}
We apply ultrafast X-ray diffraction with femtosecond temporal resolution to monitor the lattice dynamics in a thin film of multiferroic BiFeO$_3$ after above-bandgap photoexcitation. 
The sound-velocity limited evolution of the observed lattice strains indicates a quasi-instantaneous photoinduced stress which decays on a nanosecond time scale. 
This stress exhibits an inhomogeneous spatial profile evidenced by the broadening of the Bragg peak. 
These new data require substantial modification of existing models of photogenerated stresses in BiFeO$_3$: the relevant excited charge carriers must remain localized to be consistent with the data. 
\end{abstract}

\maketitle

Multiferroics have a great potential for application due to their possible coupling of ferroelectricity and magnetism \cite{eren2006a,rame2007a,mart2008a}. 
BiFeO$_3$ (BFO) is one of the few room temperature multiferroics today \cite{hill2000a,cata2009a,beli2006a,li2012a,wang2013a} and of these the only one that is a stable phase. 
Its relatively small bandgap of approx. 2.7\,eV \cite{zele2010a} renders BFO an ideal candidate for applications in spintronics and memory devices \cite{cata2009a} with a perspective for ultrafast optical switching similar to purely ferroelectric \cite{korf2007b} or magnetic materials \cite{stan2007a}.
The photovoltaic effect in this complex material and the underlying ultrafast carrier dynamics after above-bandgap femtosecond (fs) optical excitation have been studied thoroughly \cite{sheu2012a, choi2009a,yang2009a}.
{\color{black} The photoinduced currents in BFO lead to THz emission \cite{kouh2006a,rana2009a} and to a photostrictive response \cite{kund2010a}.}
All-optical experiments showed that the rapid photoinduced mechanical stress excites coherent phonons \cite{chen2012a,ruel2012a}.
The dynamics of photoinduced strains were directly and quantitatively measured in a recent synchrotron-based ultrafast X-ray diffraction (UXRD) study with a temporal resolution of 100\,ps \cite{wen2013a}.
Combined optical measurements revealed a linear dependence of the transient strain and the number of excited carriers over several nanoseconds (ns).
This led to the conclusion that depolarization field screening (DFS) including macroscopic transport of the carriers to the surface and interface could be the dominant stress generating process, although the effect of excited anti-bonding orbitals was not ruled out \cite{wen2013a}.

In this letter we report complementing UXRD experiments at a laser-driven plasma X-ray source (PXS) in order to monitor the coherent and incoherent lattice dynamics in a BFO thin film sample with sub-picosecond (ps) temporal resolution after above-bandgap excitation.
We observe a sound-velocity limited evolution of the structural response within 10\,ps indicating a quasi-instantaneous stress. 
The substantial Bragg peak broadening is a direct evidence of an inhomogeneous spatial stress profile. 
It appears quasi-instantaneously and decays on nanosecond time scales as reconfirmed by new synchrotron-based UXRD data recorded at the Advanced Photon Source (APS). 
We obtain quantitative agreement of the transient peak shift and broadening measured with both setups and can firmly conclude that the photogenerated stress driving the film expansion has a strongly inhomogeneous spatial profile in the 35\,nm thick film.
{\color{black} 
Neither the build up at sub-ps delays \cite{boja2012a} nor the remaining spatial profile at ns delays of the photoinduced stress profile in BFO can be determined by all-optical methods \cite{thom1986a} alone and hence call for the combination of the applied UXRD techniques.}

We propose a model of local charge carrier displacement within the BFO unit cells after above-bandgap excitation resulting in a lattice distortion possibly due to the inverse piezoelectric effect which drives the expansion instantaneously. 
Subsequent fast trapping of the excited charge carriers in the film maintains the stress according to the optical excitation profile over several ns until they decay radiatively \cite{sheu2012a}. 
Our experimental study provides an important benchmark for simulations of the photovoltaic response of ferroelectric oxide materials \cite{youn2012a,youn2012b}, which will have to predict strongly inhomogeneous ultrafast and long lived charge carriers.

We investigate the very same sample which was studied in Ref.~\citenum{wen2013a}.
The sample is composed of a $d=35$\,nm thick pseudocubic \hkl{0}{0}{1} BFO film epitaxially grown on a \hkl{0}{0}{1} SrTiO$_3$ (STO) substrate.
The ferroelectric polarization points along the [111] pseudocubic direction of BFO and exhibits a four-fold symmetry with most of the polarization pointing towards the surface \cite{joha2011a}.
The direct bandgap of this sample has been determined to 2.6\,eV \cite{wen2013a}. 
At the excitation wavelength of $\lambda = 400$\,nm the optical penetration $\zeta = 32$\,nm \cite{basu2008a} determines the excitation profile following Lambert-Beer's law.

The UXRD setups at the PXS and the APS have been described elsewhere \cite{zamp2009a,schi2012a,dufr2010}.
The PXS provides a temporal resolution below 200\,fs at an X-ray photon energy of 8.047\,keV (Cu K$\alpha$) and is operated in a convergence-correction mode \cite{schi2013d}. 
The X-ray and UV footprints on the sample have diameters of approx. 300\,$\mu$m and 1\,mm (FWHM), respectively.
The UV pump beam is $p$-polarized and incidents at 40$^{\circ}$ from the surface.
The synchrotron based setup provides much higher stability for long term measurements in the ns range, while excitation and probing conditions are very similar \cite{wen2013a}.

\begin{figure}
\centering
\includegraphics[width=\figurewidth]{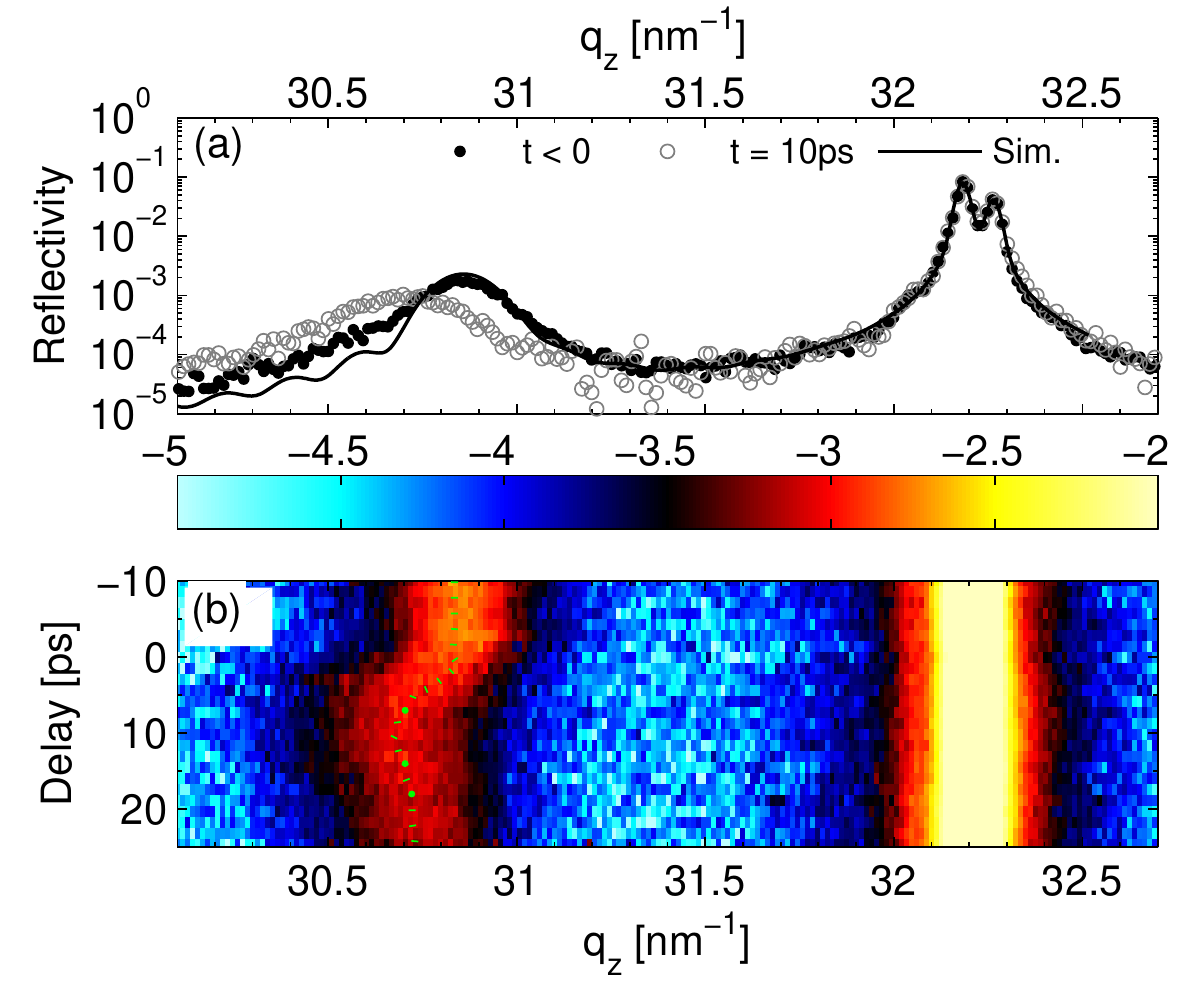}
\caption{(Color online) (a) Rocking curves of the \bragg{0}{0}{2} pseudocubic Bragg reflections from the BFO layer and STO substrate measured by the PXS.
The solid black line is a simulation of the static rocking curve.
(b) Measured transient rocking curves (diffracted intensity in logarithmic scale).
The green dashed line indicates the extracted center of the BFO peak.
}
\label{fig:rocking}
\end{figure}

Figure~\ref{fig:rocking}(a) shows the static rocking curve of the \bragg{0}{0}{2} pseudocubic Bragg peaks of the BFO and STO substrate as measured by the PXS together with a transient rocking curve at $t = 10$\,ps delay.
The high crystalline quality of the film is evidenced by the the static rocking curve which coincides with the dynamical X-ray simulation for a $d= 35$\,nm thick perfect BFO film on STO including the instrumental resolution.
Figure~\ref{fig:rocking}(b) shows the transient rocking curves for the early delays from -10\,ps to 25\,ps with the photoexcitation occurring at $t=0$.
The peak shift of the BFO \bragg{0}{0}{2} pseudocubic reflection measures the average out-of-plane strain ($z$-direction) in the layer: $-\Delta q_z(t) \propto \left<\epsilon(z,t)\right>_z $.
The observed shift to smaller $q_z$ corresponds to an ultrafast expansion $\Delta c/c \approx 0.5\,$\% of the BFO film along the surface normal without any contraction features which  were observed for the ferroelectric material PbTiO$_3$ (PTO) \cite{dara2012a}.

\begin{figure}
\centering
\includegraphics[width=\figurewidth]{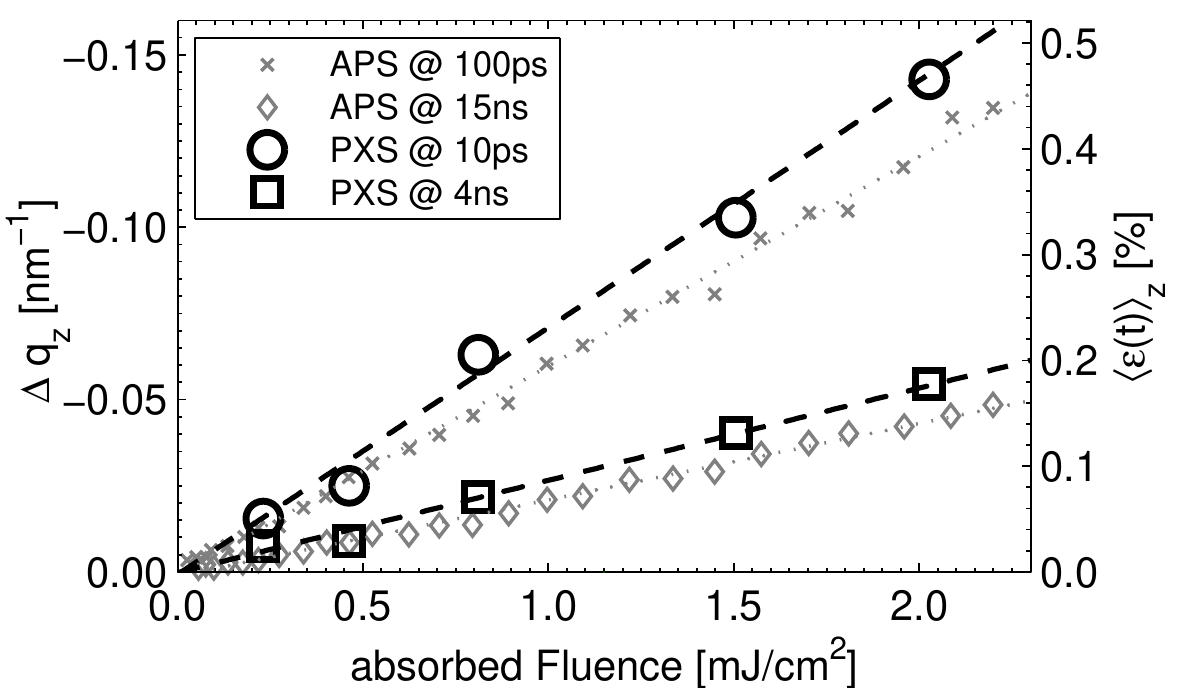}
\caption{Shift $\Delta q_z$ of the \bragg{0}{0}{2} pseudocubic BFO Bragg peak and corresponding average strain $\left<\epsilon(t)\right>_z$ measured as a function of fluence at different delays.
The dashed and dotted lines represent linear fits.
}
\label{fig:fluence}
\end{figure}

Figure~\ref{fig:fluence} shows the linear dependence of the transient peak shift on the absorbed fluence for selected delays ranging from $t = 0.01$\,ns to 15\,ns and confirms that data from the PXS and APS setups quantitatively agree within a reasonable 40\,\% recalibration of fluences between the two laboratories. 
The linear fluence dependence suggests that the origin of stress is the same for early (ps) and late (ns) delays.

\begin{figure}
\centering
\includegraphics[width=\figurewidth]{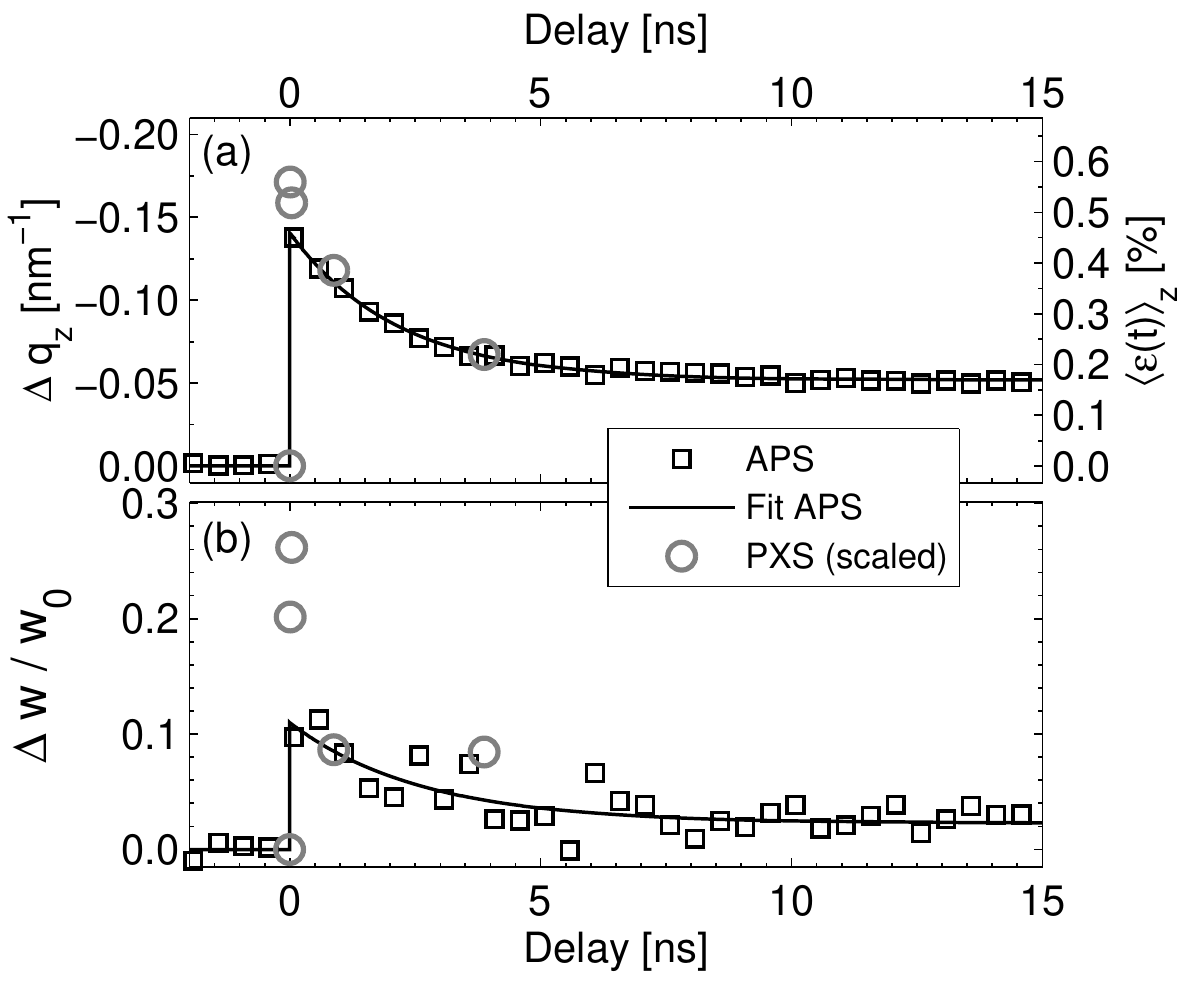}
\caption{The transient shift $\Delta q_z$ and corresponding average strain $\left<\epsilon(t)\right>_z$ (a) as well as the relative change of the width $\Delta w / w_0$ (b) of the \bragg{0}{0}{2} pseudocubic BFO Bragg peak.
The APS data was measured at an absorbed fluence of $F = 2.4$\,mJ/cm$^2$ and the PXS data was scaled according to the fluence calibration.
The black solid line represents an exponential fit of the APS data.}
\label{fig:shiftWidth}
\end{figure}

The transient BFO peak shift and width are plotted in Fig.~\ref{fig:shiftWidth}. 
The smaller peak shift of the APS data within the first 50\,ps originates from the limited temporal resolution (100\,ps) of the synchrotron-based experiments.
The decay of the shift is fitted by an exponential function with a time constant $\tau_\text{decay}^\text{shift} \approx 2.29 \pm 0.14$\,ns including an offset for very late time scales.

The width $w$ inversely depends on the number of scattering atomic layers (size broadening) and on the inhomogeneous strain-fields within the film (strain-broadening) \cite{warr1950a,will1953a}.
We can neglect mosaic-broadening for the high-quality BFO film \cite{schi2013a} and for the synchrotron-based APS experiments instrumental broadening is negligible, too.

Although the link between the peak profile and the spatial strain profile is complicated, we can assume that the change of the width $\Delta w$ depends linearly on the spatial variation $\Delta_z\epsilon(z,t)$ of the strain: $\Delta w(t) \propto \Delta_z\epsilon(z,t)$.
If the layer is homogeneously strained ($\Delta_z\epsilon = 0$) no additional peak broadening is observed ($\Delta  w = 0$).
The transient width $\Delta w$ 
reveals a significant inhomogeneous strain profile in the BFO layer over the whole observed time scale.
The exponential fit in Fig.~\ref{fig:shiftWidth}(b) results in comparable decay times for the width and shift: $\tau_\text{decay}^\text{width} \approx 2.31 \pm 0.92$\,ns.
This directly indicates that the spatial strain profile does not equilibrate within the thermal relaxation time $\tau_\text{th} = d^2\rho\, c / \kappa = 850$\,ps of the $d=35$\,nm thick BFO layer \footnote{Thermal conductivity $\kappa_\text{BFO}=3$\,W/(m\,K), density $\rho_\text{BFO}=8.34$\,g/cm, specific heat $c_\text{BFO}=0.3$\,J/(g\,K). Values taken from the supplement of Ref. \citenum{wen2013a}}.

For delays $t \gg 20$\,ps larger than the time it takes strain waves to travel twice through the thin film at the speed of longitudinal sound $v_\text{BFO} = 3.5$\,nm/ps \cite{ruel2012a}, the observed shift $\Delta q_z$ is not only proportional to the average strain  but also to the transient stress according Hook's law $-\Delta q_z \propto \left<\epsilon(z,t)\right>_z \propto \left<\sigma(z,t)\right>_z $.
The similarity of the decay times for the peak shift and broadening moreover suggests that the transient stress $\sigma(z,t)$ can be approximated by a time-invariant spatial stress profile $f(z)$ that decays in amplitude $A(t)$, i.e. no transport processes are relevant to the driving stress: $\sigma(z,t) = A(t) \times f(z)$.
We can then directly link the peak shift $\Delta q_z(t)$ to the amplitude $A(t)$ by: $-\Delta q_z(t) \propto \left< \sigma(z,t)\right>_z \propto A(t) \times \left< f(z)\right>_z$, 
as well as the peak width by: $\Delta w(t) \propto A(t) \times \Delta_z f(z)$.

\begin{figure}
\centering
\includegraphics[width=\figurewidth]{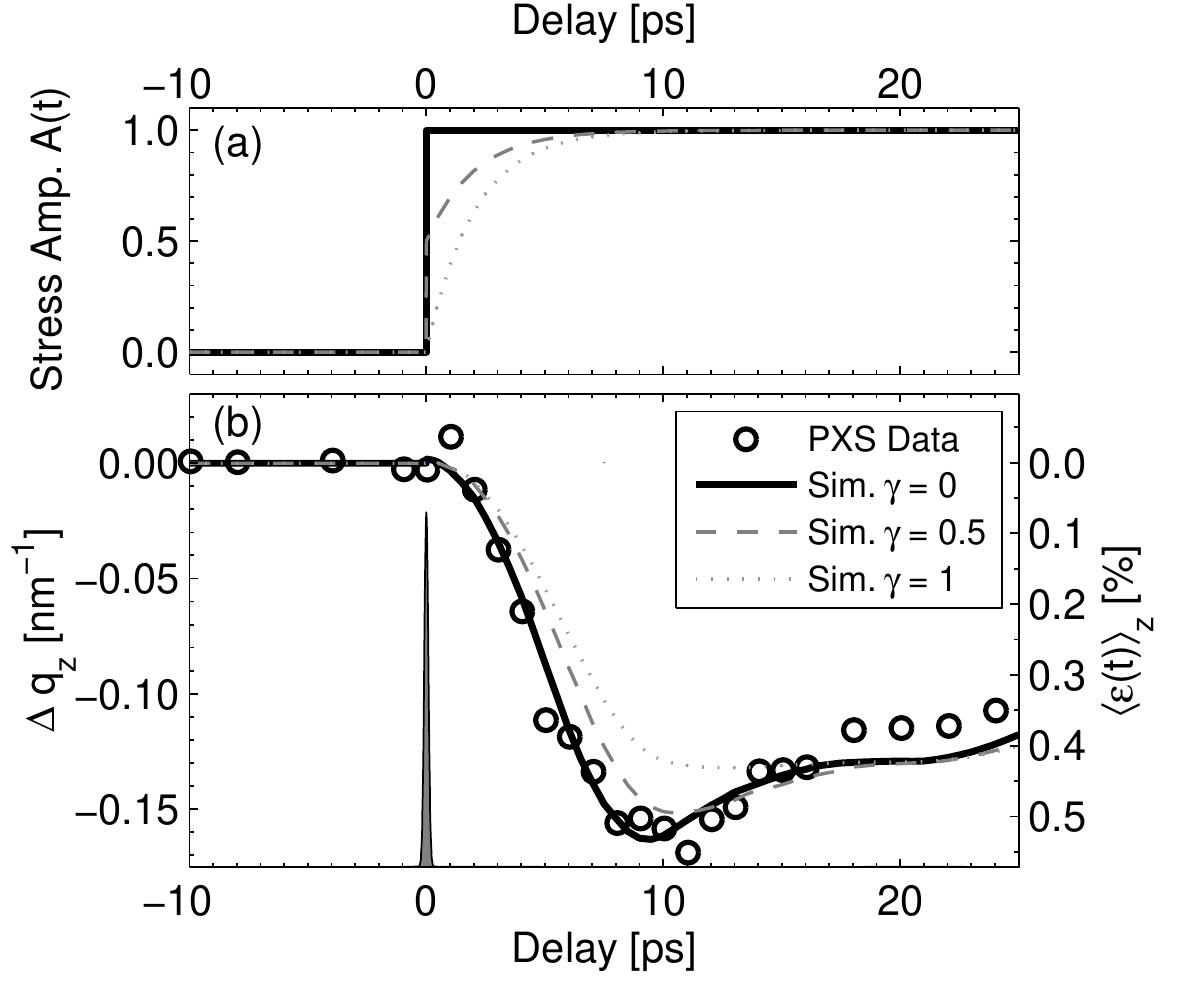}
\caption{The black circles in (b) show the transient shift of the \bragg{0}{0}{2} pseudocubic BFO Bragg peak extracted from the data shown in Fig.~\ref{fig:rocking}(b).
The black solid line represent the simulated peak shift for an instantaneous stress.
The simulated peak shift for a semi-instantaneous stress (gray dashed line) and a fully diffusive-like stress (gray dotted line) both with a time constant $\tau_\text{rise} = 2$\,ps, see text.
The corresponding stress amplitudes $A(t)$ are plotted in panel (a).
The gray Gaussian at $t=0$ in panel (b) indicates the temporal resolution of 200\,fs of the PXS setup.
}
\label{fig:shiftForce}
\end{figure}

To elucidate the origin of the photoinduced stress at very early delays ($t \leq 20$\,ps), where the stresses cannot be calculated via Hook's law, we simulate the BFO peak shift from a 1D lattice dynamics simulation \cite{herz2012b, schi2013c} of the strain profile $\epsilon(z,t)$ for the given stress
\begin{equation*}
	\sigma(z,t)\propto A(t) \times f(z) = H(t) \left(1 - \gamma\, e^{-t/\tau_\text{rise}}\right)  \times e^{-z/\zeta} \; .
	\label{eq:force}
\end{equation*}
We explicitly use the optical penetration depth $\zeta = 32$\,nm and assume a time-dependent rise $A(t)$ of the stress which includes a quasi-instantaneous stress approximated by the Heaviside step function $H(t)$ and an additional stress component growing with the time constant $\tau_\text{rise}$. 
The transient stress is plotted in Fig.~\ref{fig:shiftForce}(a) for several relative strengths $\gamma$ of the instantaneous and delayed stress components \cite{nico2011a}.
We apply dynamical X-ray theory to calculate the according rocking curves from the simulated strain $\epsilon(z,t)$ in order to extract the transient peak shift the same way as for the experimental data \cite{schi2013c}. 
We clearly obtain the best fit to the experimental peak shift for $\gamma=0$. 
In the examples shown in Fig.~\ref{fig:shiftForce} we used the time scale $\tau_\text{rise} = 2$\,ps, however, we have cross-checked this statement with additional simulations using longer timescales $\tau_\text{rise}$ for the delayed stress and also for different spatial stress profiles. In essence this proves that the dominant contribution to the stress is instantaneous and spatially inhomogeneous.

We want to emphasize the importance of several key parameters entering the simulation:

i. The temporal overlap of X-ray probe and UV pump pulses ($t = 0$) was determined independently with an accuracy of approx. $\pm 100$\,fs \cite{boja2012a}.

ii. The film thickness was determined experimentally by XRD measurements to $d = 35$\,nm.

iii. We determined the longitudinal acoustic sound velocity $v_\text{BFO} = 3.5$\,nm/ps 
independently by an oscillation of the layer thickness after photoexcitation \cite{nico2011a} (not shown) with good agreement to literature values \cite{ruel2012a}.

The DFS model proposed as origin of the fast lattice expansion in BFO \cite{wen2013a} and in ferroelectric PTO \cite{dara2012a} requires free charge carriers to travel from within the bulk material to the interfaces of the layer to screen the depolarization fields.
The ferroelectric polarization increases while the carriers drift with velocity $v_d$.
The inverse piezoelectric effect would thus yield a finite stress rise time $\tau_c$ which is required for the carriers to propagate across the film.
The typical charge carrier mobility in ferroelectrics is between $\mu = 0.1-3.0$\,cm$^2$/Vs \cite{scot2007a} and typical internal electric fields are in the range of $E \approx 200$\,kV/cm \cite{wang2003a}.
Even for a high charge carrier mobility of $\mu < 3$\,cm$^2$/Vs this leads to $\tau_c = d/v_d = d/\mu\, E > 5$\,ps for the BFO sample which contradicts the quasi-instantaneous stress that is required to drive the ultrafast lattice dynamics, c.f. Fig.~\ref{fig:shiftForce}.
We would further expect a spatially homogeneous stress profile from the DFS model due to the capacitor-like geometry in the film, which is incompatible with the significant long-lived  peak broadening shown in Fig.~\ref{fig:shiftWidth}(b).
We conclude that DFS due to diffusion of charge carriers {\color{black} across the whole film thickness} cannot be a dominant process in BFO.

Unfortunately, the fluence dependent study (Fig.~\ref{fig:fluence}) cannot differentiate between alternative stress mechanisms, since all of them essentially depend linearly on the fluence. 
In the DFS model this is only true for low fluence, since the process saturates when the depolarization field is fully screened \cite{darr1991a,babi2010a}.
For PTO \cite{dara2012a} the saturation occurs at excitation fluences of approx. 1\,mJ/cm$^2$. 
In our experiment the strain in BFO remains linear up to absorbed fluences of more than 2\,mJ/cm$^2$ (Fig.~\ref{fig:fluence}) with comparable strain levels.
At even higher fluences the BFO film starts to degrade.
Since only a very small number of the excited carriers is sufficient to drive the stress in the DFS model \cite{wen2013a}, the absence of strain saturation effect disfavors the DFS model as well, {\color{black} possibly due to the presence of a skin layer at the BFO surface which might change the ferroelectric boundary conditions \cite{mart2011a}.}

The thermal contribution to the stress can be quantitatively estimated: The electrons in the BFO film exhibit an excess energy of approx. 0.5~eV in the conduction band when they are excited with 3.1\,eV photons.
This excess energy is transferred rapidly to the lattice via electron-phonon coupling within 1\,ps \cite{wen2013a,sheu2012a} resulting in a fast temperature increase of the lattice.
We calculate this temperature jump as $\Delta T = 44$\,K averaged over the BFO film thickness for an absorbed fluence of $F = 2.4$\,mJ/cm$^2$ taking into account the internal refraction of the pump light in the BFO layer. This corresponds to a maximum thermal strain of $\epsilon = 44\,\text{K} \times 1.84 \times 10^{-5}\,\text{K}^{-1}= 0.08\,$\% which accounts only for a small fraction of the peak shift [Fig.~\ref{fig:shiftWidth}(a)]. 
A numerical simulation \cite{schi2013c} of the heat diffusion
assuming bulk values \cite{wen2013a} yields the timescale 5\,ns for reducing the average BFO temperature rise to $6$\,K.
Even the artificial introduction of a very large thermal interface resistance \cite{walk2011a} keeping the heat in the BFO layer would only explain about half of the observed shift for $t>10$\,ns. 

{\color{black}
The dominant mechanism right after excitation leads to an immediate and inhomogeneous lattice expansion, and is still important after 10\,ns.
The optical excitation transfers electrons into initially unoccupied orbitals above the bandgap where they can contribute to the lattice dynamics by deformation potential interaction \cite{zeig1992a}.
In contrast to our experimental observation this effect would lead to a contraction of the BFO film referring to Eq.~(17) of Ref.~\citenum{thom1986a} and taking the pressure-dependence of the bandgap in BFO into account \cite{gavr2008a,gome2012a}.
However, at the same time the redistribution of electron density within the BFO unit cell, also referred to as a shift current \cite{youn2012b}, influences the local ferroelectric polarization, as evidenced by THz emission \cite{kouh2006a,rana2009a}.
The change in amplitude and sign of the local ferroelectric polarization depends on the involved orbitals and thus also on the optical excitation wavelength as well as on the specific sample configuration \cite{youn2012a,youn2012b}.
For our case the photoinduced change of the local ferroelectric polarization induces a stress via the inverse piezoelectric effect which drives the ultrafast lattice expansion.
Compressive stress due to deformation potential interaction is overcompensated for the entire excited state lifetime of the charge carriers.

Theoretical and experimental studies provide evidence for the optical generation of $p$-$d$ charge transfer (CT) excitonic states in BFO \cite{pisa2009a,sheu2012a}.  
These states modify the local ferroelectric polarization and also couple strongly to the structure of the BFO unit cell \cite{pisa2009a}. 
The strong absorption edge smearing of the 2.6--2.8\,eV band \cite{basu2008a,haus2008a} and an additional weak 2.4\,eV band superimposed on its tail, measured by static spectroscopy, point to a CT instability as well as self-trapping of $p$-$d$ CT excitons and the nucleation of electron-hole droplets \cite{pisa2009a}.
Due to the rapid trapping of the CT excitonic states, the initial spatial excitation profile is maintained and the temporal dependence of the lattice strains is solely determined by electronic recombination and not by additional diffusion processes. 
We note that the optical excitation is not sufficient to drive the system into a different structural phase.
A transient structural phase, which can arise from electronic excitation or lattice distortion \cite{dieg2011a}, is not observed in our diffraction study.
}
%
%

In conclusion, we applied UXRD experiments with sub-ps temporal resolution to monitor transient lattice dynamics in a multiferroic BFO thin film after above-bandgap photoexcitation.
The peak shift reveals a rapid expansion that is only limited by the sound velocity, indicating a quasi-instantaneous photoinduced stress.
The peak broadening indicates a strongly inhomogeneous spatial stress profile for ps up to ns delays, excluding thermal stresses as the dominant process.
We propose a model of a local charge carrier displacement within the BFO unit cell after photoexcitation leading to an instantaneous stress due to the inverse piezoelectric effect. 
The fast trapping of the involved charge carriers maintains the spatial excitation profile until they decay radiatively on a ns time scale. 
We believe that this information is essential for testing theoretical models that can also distinguish the contribution of the piezoelectric effect and of the deformation potential. 
The subtle distinction between these processes could in principle be given by more nontrivial UXRD experiments where an additional (e.g. electronic) control over the polarization is implemented.

The work at Argonne was supported by the U.S Department of Energy, Office of Science, Office of Basic Energy Sciences, under Contract No. DE-AC02-06CH11357.
Work at Cornell University was supported by the Army Research Office through Agreement No. W911NF-08-2-0032.
We acknowledge the financial support for the work at Potsdam by the BMBF via grant No. 03WKP03A.

\bibliography{bibliography}

\end{document}